\documentclass[sigconf,nonacm]{acmart}
\usepackage{cleveref}
\usepackage{xspace}
\usepackage{makecell}
\usepackage{enumitem}
\usepackage{verbatim}
\usepackage{tabularx, xcolor}
\usepackage{pifont}
\usepackage{listings}
\usepackage{soul}
\usepackage{fancybox}

\newcommand{\tool}{\textsc{DataScout}\xspace}
\newcommand{\ttt}[1]{{\small \texttt{#1}}\xspace}

\newcommand{\del}[1]{}
\newcommand{\add}[1]{#1}

\newcommand{\rachel}[1]{}
\newcommand{\bhavya}[1]{}
\newcommand{\madelon}[1]{}
\newcommand{\shreya}[1]{}
\newcommand{\agp}[1]{}
\newcommand{\wenjing}[1]{}

\newcommand{\numformative}{8}

\newcommand{\numeval}{12}

\definecolor{darkblue}{HTML}{003282}
\definecolor{lightbluebg}{RGB}{245, 248, 255}

\newcommand{\focusTextBox}[2]{%
  \vspace{0.5em}
  \noindent
  {\setlength{\fboxsep}{4pt}%
  \fcolorbox{darkblue}{lightbluebg}{%
    \parbox{\dimexpr\linewidth - 2\fboxsep - 2\fboxrule}{%
      \textbf{\textcolor{darkblue}{#1}}\\
      #2
    }%
  }}
}

\newcommand{\topic}[1]{\vspace{-3.5pt}\smallskip \smallskip \noindent{\bf #1.}}

\definecolor{E1Color}{RGB}{255, 218, 185} 
\definecolor{E2Color}{RGB}{193, 225, 193} 
\definecolor{E3Color}{RGB}{173, 216, 230} 

\AtBeginDocument{%
  }

\begin{document}
\title{Rethinking Dataset Discovery with \tool{}}

\author{Rachel Lin$^{\dagger}$}
\affiliation{%
    \institution{EECS, UC Berkeley}
    \city{Berkeley}
    \state{California}
    \country{USA}
}
\email{raelin@berkeley.edu}

\author{Bhavya Chopra$^{\dagger}$}
\affiliation{%
    \institution{EECS, UC Berkeley}
    \city{Berkeley}
    \state{California}
    \country{USA}
}
\email{bhavyachopra@berkeley.edu}

\author{Wenjing Lin}
\affiliation{%
    \institution{EECS, UC Berkeley}
    \city{Berkeley}
    \state{California}
    \country{USA}
}
\email{wenjing.lin@berkeley.edu}

\author{Shreya Shankar}
\affiliation{%
    \institution{EECS, UC Berkeley}
    \city{Berkeley}
    \state{California}
    \country{USA}
}
\email{shreyashankar@berkeley.edu}

\author{Madelon Hulsebos}
\affiliation{%
    \institution{Centrum Wiskunde \& Informatica}
    \city{Amsterdam}
    \state{}
    \country{Netherlands}
}
\email{madelon@cwi.nl}

\author{Aditya G. Parameswaran}
\affiliation{%
    \institution{EECS, UC Berkeley}
    \city{Berkeley}
    \state{California}
    \country{USA}
}
\email{adityagp@berkeley.edu}

\renewcommand{\shortauthors}{Lin and Chopra et al.}

\newif\ifanonymous
\anonymousfalse 
\ifanonymous
\else
  \thanks{$^\dagger$Co-first authors. Corresponding author: Bhavya Chopra.}
\fi

\begin{abstract}

{\em Dataset Search}---the process of finding appropriate datasets for a given task---remains a critical yet under-explored challenge in data science workflows. Assessing dataset suitability for a task (e.g., training a classification model) is a multi-pronged affair that involves understanding: 
data characteristics (e.g. granularity, attributes, size), 
semantics (e.g., data semantics, creation goals), 
and relevance to the task at hand. 
Present-day dataset search 
interfaces are restrictive---users struggle 
to convey implicit preferences and lack visibility 
into the search space 
and result inclusion criteria---making 
query iteration challenging.  
To bridge these gaps, we introduce \tool to
proactively steer users through the process of dataset discovery via---{\em (i)} 
AI-assisted query reformulations 
informed by the underlying search space, 
{\em (ii)} semantic search and filtering 
based on dataset content, including attributes (columns) 
and granularity (rows), and {\em (iii)} 
dataset relevance indicators, generated dynamically 
based on the user-specified task. 
A within-subjects study with \numeval{} 
participants comparing \tool{} to keyword and semantic dataset search 
reveals that users uniquely employ \tool's features not only for structured explorations, but also to glean feedback on their search queries and build conceptual models of the search space. 
\end{abstract}



\maketitle

\section{Introduction}

Finding the right dataset, given a data analysis or machine learning task, is one of the most challenging problems for data scientists and analysts today~\cite{chapman2020survey}. This problem of {\em dataset search} is only growing more urgent---with
organizations often accumulating
tens of thousands of tables in their data lakes~\cite{kayali2024mind}. Dataset search is difficult for a couple of reasons. First, real-world data is inherently messy: tables vary widely in quality and metadata completeness, with many lacking proper descriptions, having ambiguous column names, or containing outdated information~\cite{sostek2024discovering}. Second, users rarely know exactly what they are looking for~\cite{hulsebos2024took}. They might have a general task in mind, like training a machine learning model to predict some phenomenon, but do not know which datasets would be compatible with their task. 

Recent advances in Large Language Models (LLMs) have demonstrated the potential to address some of the aforementioned challenges. Embedding models enable us to transform unstructured text into numerical representations (i.e., embeddings) that capture semantics, allowing systems to perform a {\em semantic search} to find relevant datasets, even when the exact terminology differs~\cite{zhao2024user}. For example, Olio~\cite{setlur2023olio} can interpret a natural language (NL) question like ``how has unemployment changed since 2020'' and find relevant datasets---even
if the metadata does not have a perfect
keyword match with the question. However, semantic search 
of this form is often opaque to users, making it difficult to understand {\em why} a particular dataset appears in the search results, or {\em how} it relates to their query---plus users are unable to adaptively
explore the content within datasets, including the columns/attributes, and temporal/spatial granularity. Overall, despite these advances in interpreting NL queries, present-day dataset search interfaces---be it semantic or keyword-based---provide limited support for search expressiveness---illustrating a wide gap between what technology can enable, and what interfaces currently facilitate. 

\begin{figure*}
    \centering
    \includegraphics[width=0.95\linewidth]{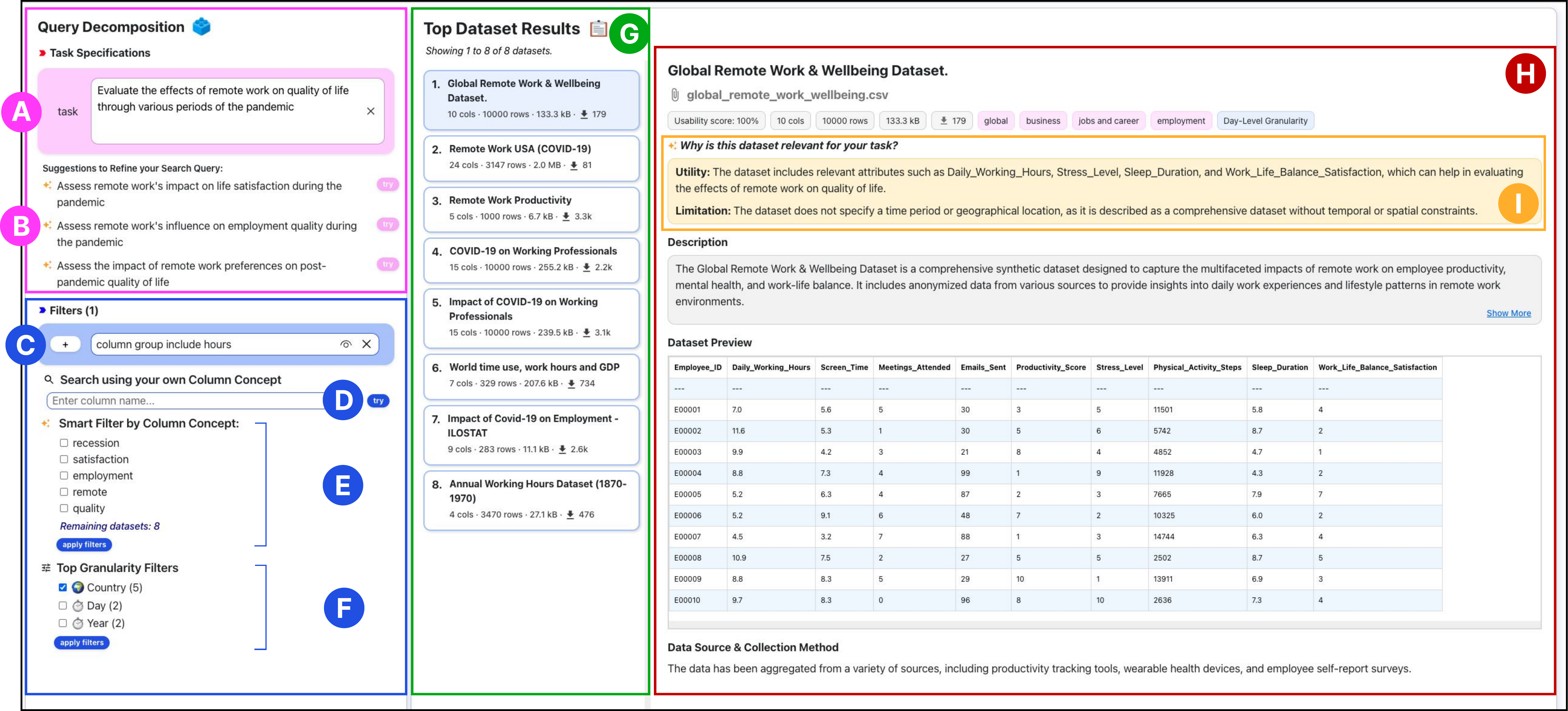}
    \caption{\tool{}---a proactive dataset discovery interface. (A) Users begin by specifying their query as keywords, phrases, or complete sentences. (B) \tool provides query reformulation suggestions to bridge the gap between the user's query and datasets available in the search space. (C) Users may add exact matching-based or semantic filters, (D) search by attribute, apply (E) suggested attribute filters, or (F) suggested temporal and spatial granularity filters. (G) Users can explore ranked dataset search results in a consolidated view. (H) Selecting a dataset reveals its metadata, tags, description, preview, collection details, and (I) task-specific relevance indicators generated on-the-fly, highlighting utilities and limitations of the dataset.}
    \label{fig:interface}
\end{figure*}

Moreover, users typically lack awareness of available datasets, and must learn about the dataset landscape through the search results themselves,
which subsequently inform refinements of their queries. This makes dataset search an inherently exploratory, iterative, and often tedious process requiring multiple query reformulations and result assessments~\cite{hulsebos2024took}. 
Users have to rely on the assistance
of colleagues for starting points,
or even direct identification of the relevant
datasets---indicating just how poor
present-day dataset search interfaces are in supporting iterative exploration.

\noindent In this work, we explore the design of dataset search systems that can proactively support users' iterative discovery process. To do so, we first conducted a formative study to identify aspects of users' dataset search workflows that could be amenable to automated assistance. 
Our study findings reveal that:
\begin{itemize}[leftmargin=*]
\item Users \textit{\textbf{lack efficient means to express intent}}, finding dataset search interfaces to be restrictive in filtering based on content, attributes (columns) and granularity (rows);
\item Users receive \textit{\textbf{limited insight into characteristics of the dataset search space}}, such as the space of possible 
columns across the returned results; 
\item Users \textit{\textbf{struggle to reformulate their queries}} when encountering overly selective or irrelevant datasets; and
\item Users \textit{\textbf{spend significant time in assessing dataset relevance}} in context of their analytical needs, especially when the dataset description focuses on what the dataset contains, not the purposes it can be used for.
\end{itemize}
These limitations underscore the need for dataset search interfaces that {\em proactively empower users with feedback and assistance to iteratively reformulate queries, interpret search results, and navigate the dataset search space.}
We present \tool{}, a dataset search tool that proactively steers users through the process of dataset discovery (\Cref{fig:interface}). \tool{} assists users in finding target datasets by being cognizant of both the user-specified task as well as the underlying space of results. \tool{} offers three key LLM-powered semantic assistance features: {\em (i)} \textbf{proactive query reformulation (\Cref{fig:interface}B)} to bridge the gap between users' search queries and the underlying search space, \add{ensuring that each reformulation is both diverse and grounded in actual search results by covering a subset of the dataset corpus,} 
{\em (ii)} \textbf{semantic search (\Cref{fig:interface}D)} and \textbf{filtering} based on dataset content, including \textbf{attributes (\Cref{fig:interface}E)} and \textbf{granularity (\Cref{fig:interface}F)} to help users appropriately narrow down the search space, and {\em (iii)} \textbf{semantic relevance indicators (\Cref{fig:interface}I)} generated on-the-fly based on the user-specified task to help them assess dataset relevance rapidly.

To enable these interactions, we split \tool{}'s workflow across offline and online components---balancing a trade-off between semantic expressiveness and latency. We precompute embeddings, indexes, and inferred metadata where possible (e.g., for semantic dataset and attribute searches), while relying on LLMs-in-the-loop for dynamic features requiring search context (i.e., generating query reformulations, semantic filtering suggestions, and task-specific relevance indicators). This hybrid architecture allows \tool{} to deliver rich, personalized assistance without prohibitive latency, reflecting a broader systems-level challenge of designing intelligent interfaces that combine responsiveness with semantic assistance.

To evaluate \tool{}, we conducted a within-subjects study with \numeval{} participants; comparing its semantic reformulation, filtering, and relevance assessment modalities with traditional keyword and semantic search interfaces (\Cref{sec:user-study}). We find that users leveraged \tool's features not only for more structured and intentional navigation of the dataset search space, but also as implicit feedback mechanisms---helping them reflect on their queries, make sense of individual datasets, and better understand the overall search landscape.
Overall, we make the following contributions:
\begin{itemize}[leftmargin=*]
\item Design considerations for semantic dataset discovery interfaces, derived from prior work and our formative study ($n=\numformative$);
\item Design and implementation of \tool{}, a dataset discovery tool to proactively steer users towards desirable datasets; and 
\item Empirical findings from a within-subjects user study ($n=\numeval$) demonstrating how users uniquely leverage \tool's suggestions and assistance for sensemaking.
\end{itemize}
\section{Related Work}
\tool builds on research in information seeking theories, dataset discovery interfaces, and web search tools. 

\topic{Information Seeking Models and Interfaces}
Information seeking 
has a long history of theories
and successful interfaces~\cite{hearst2009search}.
Traditional information
seeking theories
describe iterative cycles of query specification,
examination of results, and 
reformulation, until the need is satisfied~\cite{marchionini2006exploratory, shneiderman1987designing}.
Other classical models
conceptualized this as information foraging~\cite{pirolli1999information},
where users follow ``information scents'' across content ``patches.''
This framework was then extended
to encompass a subsequent stage of ``sensemaking,''
the process of synthesizing and contextualizing information~\cite{pirolli2007IFT}. Sensemaking helps users understand what they are finding along the way and contextualize it with their own objectives~\cite{RUSSELLROSE201323, bates1989design}. \del{Bates' berry-picking model, further emphasizes the benefits of information seeking as a series of learnings along the way, as opposed to just meeting a desired information seeking target.}
In our context, these models underscore the intertwined nature of exploration and sensemaking---where users refine goals and progressively discover dataset characteristics ``along the way.'' 
Ideal dataset discovery systems must guide users to: {\em (i)} formulate their query to narrow down to the correct subset of the search space, and {\em (ii)} contextualize the surfaced search results with their analytical intents and assess their relevance.
These information seeking models 
have notably shaped web search systems.
Modern web search interfaces support keyword search, auto-suggestions, 
related query suggestions, and empower users
to filter
results based on attributes and facets
like time and file type~\cite{lee2009facetlens, smith2006facetmap, tunkelang2022faceted, lee2005understanding}.
\del{In domains like e-commerce or travel, 
search interfaces enable search across orthogonal categories (e.g. airline, departure time, number of stops), called facets, that are predefined by 
the interface designers.
Faceting has been studied in various search contexts.}

Recent work on web search 
and information retrieval
continues to build on these foundations. \citet{palani2021conotate} show that users' objectives evolve through inspecting search results, particularly as they gather more information about a new problem area with ill-defined information seeking goals. 
This is 
relevant in dataset search,
since users may still be learning domain-specific vocabulary and assessing possibilities in early stages---as opposed to knowing precise datasets of interest upfront.
Tools like Sensecape and CoNotate also provide suggestions for web search queries grounded in the user's context to close information gaps~\cite{suh2023sensecape, palani2021conotate},
while 
other recent work explores how to best support the sensemaking process in a lightweight in-context manner~\cite{kuznetsov2022fuse, kuznetsov2024tasks, morris2010wesearch}. \add{Luminate uses an LLM to generate structured ``dimensions'' of design spaces for creative exploration~\cite{suh2024luminate}. While these papers show the value of LLM-driven reformulations, unconstrained reformulations can derail the dataset search experience and erode user confidence by yielding queries that have no matching datasets. Instead, \tool{}'s reformulations are informed by search results: ensuring that each reformulation covers a subset of the results, and is diverse---thereby being grounded in actual dataset availability.}

LLM-generated relevance indicators have also shown promise. \citet{liu2024selenite} find that users benefit when systems surface decision-relevant cues aligned with criteria previously found helpful for decision-making. \tool{} extends this idea by generating dynamic, query-specific dataset relevance indicators.~\citet{koesten2021talking} identify key dimensions users assess during search for dataset suitability. 
These include data distributions, granularity, quality, possible questions the data can answer, and creation details. \tool{} surfaces relevance cues aligned with these dimensions to support dataset sensemaking.

In recent years, conversational search has emerged as a new search paradigm, leveraging clarifying questions as mixed-initiative probes to iteratively refine user intent~\cite{vtyurina2017exploring, radlinski2017theoretical, zhang2018towards, rosset2020leading, mo2024survey}. This paradigm has been adapted by dataset search tools like Olio~\cite{setlur2023olio} and MetaM~\cite{galhotra2023metam}. However, these methods still rely on users to identify and formulate their dataset requirements as queries or questions, providing limited proactive guidance to them.

\topic{Dataset Search: Challenges and Recommendations}
\label{related-dataset-search-qualitative}
Dataset search poses unique challenges, distinct from traditional web search.
Users span a range of expertise and goals,
where in many cases the goals (e.g., training a machine learning model) are far removed
from the datasets. 
The datasets themselves
are often hard to peruse manually.
Despite advances in interpreting
natural language intents, users still struggle with incomplete and inconsistent metadata~\cite{fan2023table, sostek2024discovering}, expressing information seeking needs as structured search constraints~\cite{koesten2017trials}, and assessing dataset relevance~\cite{koesten2020everything, sostek2024discovering}. These challenges lead users to face gulfs of execution (difficulty articulating intents to dataset search interfaces) and evaluation (difficulty interpreting if the system perceived their search intent, and if it is reflected by the surfaced datasets)~\cite{norman2002design}.

A recent survey by~\citet{hulsebos2024took} further highlights how data practitioners rely on trial-and-error search refinements to overcome these barriers, calling for interfaces that better support iterative search refinement and focus on users' analytical goals. 
Recent work also emphasizes the need for better query assistance, dynamic metadata filters, and clearer descriptions to aid sensemaking~\cite{zhao2024user}. We build on these papers by conducting a formative study (\Cref{sec:formative}) that directly observes users' search workflows in modern dataset search interfaces to 
identify pain points and inform the design of \tool.

\topic{Dataset Search: Mechanisms and Objectives}
Popular dataset search tools 
employ various approaches to retrieve relevant datasets, in both the input dataset space and the underlying search mechanisms. 
Repositories such as Kaggle and
HuggingFace
support keyword-based search over dataset descriptions.
Others use semantic approaches---for example, Google Dataset Search indexes datasets from repositories and individual web pages, and uses semantic matching~\cite{brickley2019google, sostek2024discovering}. Databricks Search and Snowflake Universal Search combine keyword and semantic search~\cite{Tarakad2024, Zhang2024}. 
However, these systems typically offer static metadata filters, lack support for
iterative exploration by helping users
reformulate questions,
and provide no cues for {\em why}
a given dataset matches a query.

Dataset search spans two separate stages~\cite{chapman2020survey}: {\em (i)} task-based dataset search---finding an initial dataset for a given task; and {\em (ii)} join/union dataset search---enriching an already-identified dataset via dataset joins or unions. 
The former is driven by keyword or semantic queries, while the latter uses an input table targeted for enrichment.
For task-based search, recent efforts focus on scalability, privacy, and efficiency~\cite{herzig2021open, bogatu2022voyager, castelo2021auctus, fernandez2018aurum}. For join/union dataset search, recent efforts identify semantically equivalent attributes for ``joins'', or aligned schemas for ``unions'' to enrich the previously identified dataset~\cite{khatiwada2023santos, esmailoghli2023blend, fan2022semantics, leventidis2024large, bogatu2022voyager, huang2023fast, galhotra2023metam}, but do not focus on interface design. 
\del{For instance, Metam [14] employs heuristics to recommend join/union datasets that optimize user-specified analytical goals (such as improving prediction accuracy). These approaches have not yet focused on interfaces for join/union search.}
Overall, qualitative findings from multiple studies highlight that task-based dataset search remains largely unsupported~\cite{hulsebos2024took, koesten2017trials}. 
With \tool{}, we aim to address this gap by exploring proactive interfaces for task-based search.  

Perhaps most closely related to our work
is Olio~\cite{setlur2023olio}, a semantic question-answering system that surfaces datasets by combining natural language queries with dynamically generated and pre-authored visualizations. Olio enhances exploratory search by letting users scan visualizations to assess dataset relevance.\del{While Olio interprets users' natural language queries to surface datasets,} We build on the semantic dataset search approach adopted by Olio and redirect our focus on iterative---and proactive---query refinement: guiding users to progressively explore the search space as they learn about the underlying data. 
Unlike Olio, which assumes a predefined question for which
a visualization exists in the data, we support the iterative process of discovering the search space and task requirements.
\del{Olio, for example, does not consider filtering
based on dataset contents (such as attributes and granularity).}

\section{Design Considerations for \tool{}}
To identify dataset discovery workflows that could benefit from assistance with the challenges noted in prior work (\Cref{related-dataset-search-qualitative}), we conducted a formative study with \numformative{} participants (F1–F\numformative), and identified four design considerations (DC1–DC4) for \tool{}.

\subsection{Formative Study}
\label{sec:formative}
Participants were recruited via: {\em (i)} contacting a mailing list of data science professionals maintained by our research group, {\em (ii)} messaging on Slack and Discord channels with data science, ML, and AI graduate students, and {\em (iii)} posting to X. All participants voluntarily participated in the study and agreed to have their screen-sharing sessions recorded for transcription and analysis. Table \ref{table:formative-participants} reports participant background and formative study tasks. 

Participants took part in a 40-minute contextual inquiry session via Zoom. We began with a round of introductions, and observed participants perform a dataset search task of their choice with any preferred tool(s) (\Cref{table:formative-participants}), as they thought-out-loud about their actions. We concluded by asking clarifying questions and gathering open-ended feedback on their dataset search experiences. This study received approval from our Institutional Review Board (IRB).

We analyzed transcripts supplemented with notes documenting participant actions. Two authors performed reflexive thematic analysis through open coding of the transcripts, notes, and screen recordings, followed by identifying axial codes~\cite{braun2006using, braun2019reflecting}. The authors subsequently performed a second iteration to refine themes and motivate design considerations for \tool{}.

\subsection{Findings}
Here, we present our findings, identifying challenges in how users express and reformulate their dataset search intents, while attempting to assess dataset suitability and the underlying dataset landscape. We further highlight design considerations (DCs) stemming from these insights in-situ.

\begin{table}[t]
\caption{Formative study of participants' backgrounds, tasks, and choice of platforms.}
\label{table:formative-participants}
\small
\scalebox{0.76}{
\begin{tabular}{p{0.3cm} p{2cm} p{3.8cm} p{3.8cm}}
\toprule
\textbf{ID} & \textbf{Background} & \textbf{Task} & \textbf{Platform(s)\textsuperscript{1}} \\
\midrule
F1 & HCI, AI Research & Collections of web-service URLs & Perplexity, Google Dataset Search \\
F2 & ML Engineer & Game actions data for emulations & HuggingFace \\
F3 & Data Analyst & Pharmaceutical drug marketing & Kaggle, Google Dataset Search \\
F4 & Art \& technology & Art History and Provenance data & Kaggle, Artsy Genome \\
F5 & ML Engineering & Populating a data lake & Kaggle \\
F6 & Bioinformatics & RNA Sequences for Epilepsy & GEO, Google Dataset Search \\
F7 & AI Code-Gen & Code performance benchmarks & Papers with Code \\
F8 & Marine Science & Land use for Clean Energy & Census Data \\
\bottomrule
\end{tabular}}
\begin{flushleft}
\textsuperscript{1}\textit{Platforms spanned semantic-based (Perplexity, Google Dataset Search), keyword-based (Kaggle, GEO, Census Data, HuggingFace, Artsy Genome), and hybrid (Papers with Code) dataset search mechanisms.}
\end{flushleft}
\end{table}

\topic{Users do not express search criteria due to the fear of missing out on potentially-relevant datasets} 
Participants had several implicit relevance criteria which were not specified to dataset search platforms. For instance, when looking for datasets to train a classifier on misinformation, F4 wanted their dataset to have as many features (columns) as possible, and while looking for a collection of URLs of web-services belonging to varied economic sectors, F1 wanted the dataset to have at-least 1000 rows. On the other hand, when F1 switched from using Google Dataset Search to Perplexity, they explicitly mentioned their preference for ``1000+ rows'' in their prompt.  While such criteria could be specified as filters, participants preferred to keep their search open-ended to avoid filtering out potentially useful datasets. 

\focusTextBox{(DC1) Expression of Free-Form Intent}{
Enable users to express varied facets of their analytical and dataset search
intents in as much detail as desired, without significantly constraining the
volume of dataset search results.
}

\topic{Users desire dataset content-based filtering after initial rounds of sensemaking} Several participants wanted to filter datasets based on their content (F1, F4--F6, F8), that {\em ``simply cannot be specified to the interface''} (F2). Filtering based on content such as attributes (columns) and data granularity (rows) is not supported by present-day dataset search interfaces, as also identified by \citet{hulsebos2024took}. F5 mentioned that even if the system did support searching or filtering by column names, they would run into a {\em ``schema misalignment''} problem, defining it as {\em ``datasets using different vocabulary to refer to the same concepts,''} and elaborated using an example from movie datasets---{\em`` datasets can have different column names for the movie title, such as `title', `movie name', or `movie title,' making it impossible to apply filters.''} F3 and F8 wanted to filter datasets based on data granularity, e.g., drug-specific sales records, as opposed to pharmaceutical brand-level sales for F3; and latitude/longitude-level spatial resolution, as opposed to region names for F8.

Further, participants incrementally developed an understanding for desirable attributes they wanted to be present in their data as they inspected dataset search results, echoing the findings of~\citet{palani2021active}. For instance, after looking through top search results for LLM-code generation benchmark datasets, F7 realized that most datasets do not contain the prompt provided to the LLM to generate code, and expressed the need have the ``prompt'' column in all dataset results. F4 articulated this as an instance of {\em ``recognition over recall,''} i.e., having to recognize the need for specific attributes or data granularity after initial sensemaking of search results---as opposed to consciously acknowledging them from the get-go.

\focusTextBox{(DC2) Semantic Dataset Content-based Filtering}{
Provide users the agency to identify and place fine-grained attribute (column) and granularity (row) semantic filters at the dataset content level, rather than just the dataset description.}

\topic{Lack of query-specific dataset relevance indicators slows-down dataset discovery}
Traditional dataset search tools failed to offer indications of relevance to the query beyond the dataset title and preview, number of downloads, and column distribution histograms to users. Some participants vocalized challenges with having to read long data descriptions to identify any caveats, and oftentimes realized critical limitations of the data after having downloaded it and spent significant amounts of time to perform exploratory data analysis (EDA) (F1, F2, F5--F8). In contrast, we observed F1 using Perplexity\footnote{an AI-powered search engine and chatbot: \url{https://www.perplexity.ai/}} to enlist dataset sources along with contextualized explanations for how a given dataset might fit their needs---helping them assess dataset suitability.  
Additionally, multiple participants frequently questioned why the surfaced datasets in the search results were relevant to their search query, especially for semantic search engines like Google Dataset Search (F1, F3, F4, F6, F8). F8 brought up feedback mechanisms provided by Google's traditional web search, such as the bold-font highlighting of matched terms---helping them infer how the search result is relevant to their query---and pointed out their absence in dataset search tools. 

\focusTextBox{(DC3) Dataset Suitability Assessment}
{Facilitate sensemaking of dataset relevance and result inclusion criteria in
context of the user-specified search query and filters.}

\topic{Irrelevant or overly selective dataset search results halt query iteration}
As users of semantic dataset search systems lacked transparency on dataset inclusion criteria, they were frequently confused by irrelevant search results, blocking them from iterating over or reformulating their query (F1, F3, F6, F7). On the other hand, users of keyword-search platforms expressed frustration with overly selective search results (F2, F3, F4, F8). 

For instance, F4's search query to look for ``historical artworks with images'' yielded only 4 search results, none of which were related to art history. In such cases, participants engaged in the well documented trial-and-error query reformulation workflows to widen their scope~\cite{marchionini2006exploratory}---while still failing to identify relevant datasets. Prior work has also identified how gauging the dataset search space is overwhelming for users~\cite{hulsebos2024took}.  

\focusTextBox{(DC4) Guide Query Reformulation}{
Bridge the gap between search queries and underlying dataset landscape to
overcome overly selective or irrelevant results.}

We shaped \tool{} with the derived design considerations (DC1--DC4). In the following section, we present a walkthrough of \tool's key features and capabilities. 
\begin{figure*}
    \centering \includegraphics[width=0.88\linewidth]{figures/walkthrough.pdf}
    \caption{Walkthrough of \tool{}.
    Dana expresses her intent using the (A) getting started card. \tool{} retrieves results. Dana reviews (B) query reformulation suggestions and hovers to view explanations. She clicks on the third suggestion---refreshing the results. Dana uses semantic (C) attribute and (D) granularity filter suggestions to narrow her search to datasets containing logged employee hours and country-level data. She inspects dataset relevance using (E) dynamic task-specific relevance indicators, and (F) dataset description summaries.}
    \label{fig:walkthrough}
\end{figure*}

\section{Walkthrough of \tool{}}
\label{sec:scenario}

Here, we provide a walkthrough of \tool{} with
Dana, a journalist, who has been inspecting the world happiness reports spanning 2015--2025.\footnote{The World Happiness Report is an annual publication that ranks countries based on how happy their citizens perceive themselves to be. URL:~\url{https://worldhappiness.report/}}
She wishes to observe the impact of fine-grained lifestyle changes on the reported aggregate happiness scores. To do so, Dana decides to focus on datasets overlapping with the COVID-19 pandemic---to observe the impact of stark differences in lifestyles (e.g., confinement, reduced physical activity, and remote work and education) on happiness scores. 

Dana now turns to \tool to search for datasets. Since this is a new area of exploration for her,
she begins by using the {\em \textbf{Getting Started card}} (\Cref{fig:walkthrough}A), where she specifies her intent as a regression analysis task, while expressing her query in natural language as ``datasets indicating quality of life before, during, and after the COVID-19 pandemic'' (supporting \textbf{DC1}). 
In response, \tool{} surfaces search results and proactively inspects them to identify pertinent themes. For Dana's query, \tool{} learns that the search results spanned shifts in inflation, social media trends, and employment patterns. \tool{} then uses these insights to propose three {\em \textbf{query reformulation suggestions}} (\Cref{fig:walkthrough}B) centered around Dana's task, in an attempt to bridge the gap between her query and the underlying dataset search space (supporting \textbf{DC4}). The suggestions help Dana by providing her inspiration for analytical directions she can pick. She hovers over each suggestion to inspect explanations for the suggested queries, and the number of datasets matching the theme. She selects the suggestion: ``analyze the impact of the pandemic on remote work and work-life balance,'' since it is an evident indicator of happiness owing to sudden transformations in work patterns during the pandemic. \tool{} refreshes the search results.

As Dana inspects the datasets, she realizes the need for three additional requirements. 
First, since Dana mentioned the pandemic in her query, \tool's {\em \textbf{task-specific relevance indicators}} (\Cref{fig:walkthrough}E) surface the data collection time-period for each dataset she explores. This reminds her to look for datasets where the time-range of data collection overlaps with the 2015--2025 year bracket. \tool's semantic relevance indicators allow her to quickly glean this information, helping her efficiently identify data sources that align with her intent (supporting \textbf{DC3}).

Second, \tool{} inspects all search results and identifies attributes most relevant to Dana's query---surfacing them as {\em \textbf{semantic column concept filters}} (\Cref{fig:walkthrough}C). Observing suggestions for `hours,' `vacations,' and `stress' help Dana realize that she wants to have these attributes in her target dataset. To only focus on datasets with quantitative measures like logged work hours, Dana applies the semantic column concept filter to narrow down the results (supporting \textbf{DC2}). Third, as she continues to inspect datasets, she realizes that to make meaningful comparisons with the world happiness reports, she needs the geographical granularity of her data to be country-level. To do so, she uses \tool's {\em \textbf{semantic geo-granularity filter}} (\Cref{fig:walkthrough}D), setting ``country'' as the data granularity level (supporting \textbf{DC2}). Dana applies these filters and continues to iteratively evaluate dataset suitability.
\section{\tool{}: System Implementation}

\tool{} is implemented as a web-based application using React and TypeScript for the frontend, with a backend powered by Python, Flask, and a PostgreSQL database of datasets fetched from Kaggle, detailed in the next section. In addition to \tool's features that proactively support and aid semantic dataset search, 
it also includes a few standard features found in Kaggle and Google Dataset Search, including: ranking of datasets based on semantic relevance; dataset pages with metadata, description, and a preview; and filters over size, shape, title, description, and tags/keywords.

\begin{figure*}
    \centering \includegraphics[width=0.9\linewidth]{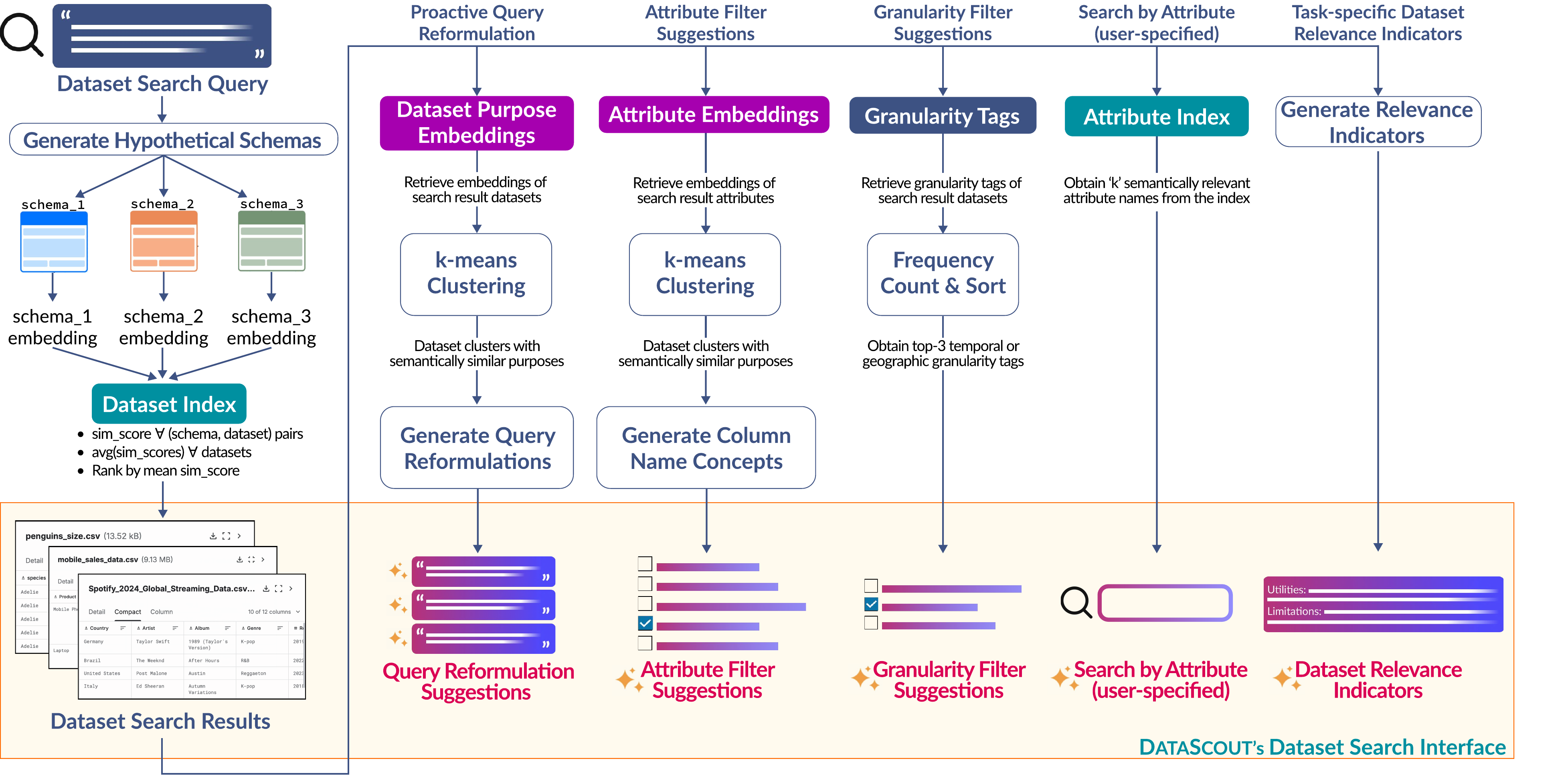}
    \caption{Online dataset search assistance. The user query is used to generate hypothetical schemas to retrieve matching datasets from the Dataset Index (\Cref{sec:system-impl-data-collection}). \tool{} proactively generates query reformulations, semantic attribute and granularity filter suggestions, and dataset relevance indicators---grounded in dataset search results and the search query. Users may accept a reformulation, apply filters, search by attributes, or inspect relevance indicators.}
    \label{fig:online-workflow}
\end{figure*}

\tool{}'s design distributes the workload across offline and online stages of interaction. Offline, we precompute embedding collections (i.e., compressed semantic representations) and build indexes for dataset and attribute (or column) search (\Cref{fig:framework}). Then, online, to enable contextualized assistance grounded in the user's search query and surfaced dataset search results, \tool{} relies on LLM-in-the-loop workflows (\Cref{fig:online-workflow})---generating: {\em (i)} query reformulation suggestions; {\em (ii)} semantic data content-based attribute and granularity filter suggestions; and {\em (iii)} dataset relevance indicators. This hybrid architecture enables \tool{} to avoid prohibitive latencies, while still providing in-situ and personalized assistance. In the following subsections, we detail our offline data collection and indexing stages, and online feature-specific implementation details.

\subsection{Offline Data Collection and Indexing}
\label{sec:system-impl-data-collection}

\begin{figure*}
    \centering
    \includegraphics[width=0.8\linewidth]{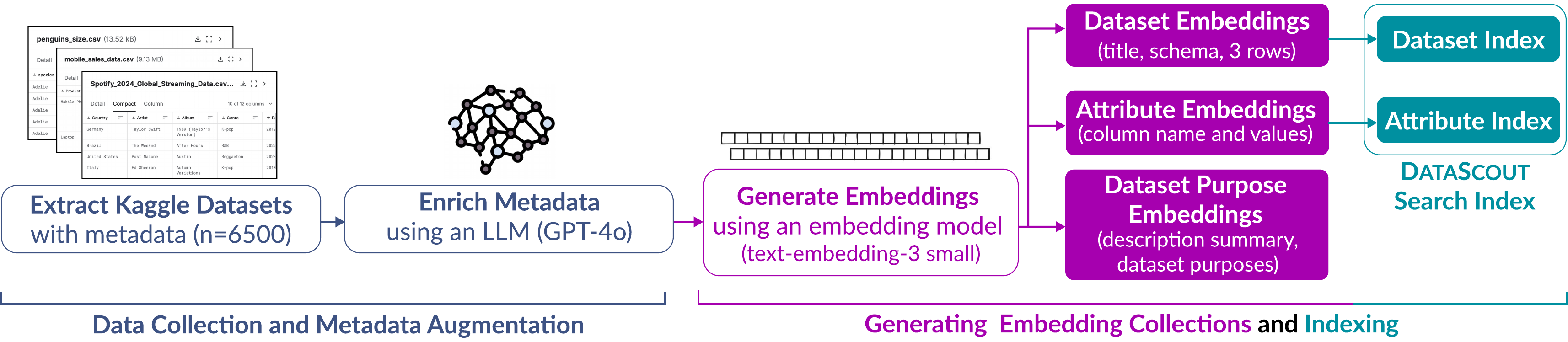}
    \caption{Offline dataset collection, augmentation, embedding generation and indexing for \tool{}. 
    }
    \label{fig:framework}
\end{figure*}

\Cref{fig:framework} and ~\Cref{tab:offline-computations} provide an overview of our data collection, preprocessing and indexing pipeline. We collected datasets from Kaggle using their API, obtaining over 6,500 unique tables (belonging to over 3150 datasets---where each dataset contained one or more tables within).
For each table, we extracted metadata, including: title, filename, description, 
tags, dataset size, number of rows and columns, usability score, number of downloads, and a sample of 10 rows with headers, formatted as a markdown table. 
To standardize and enrich the available metadata, we used OpenAI's \ttt{gpt-4o-mini} model to generate: {\em (i)} concise one-line dataset summaries using descriptions from Kaggle (\textbf{DC3}), {\em (ii)} column descriptions and inferred data types (\textbf{DC3}), {\em (iii)} data source and collection methods (\textbf{DC3}), {\em (iv)} temporal and spatial granularity by looking at example rows (\textbf{DC2}), and {\em (v)} the set of purposes or use-cases the dataset might support (e.g., regression, classification, visualization, or temporal analysis) (\textbf{DC3, DC4}). The prompts to generate these additional dataset metadata are in~\Cref{appendix:prompts-offline}. 

Then, to support the previously identified design considerations, we generated three different sets of embeddings\footnote{Embeddings are compressed vector representations of the data; with similarity of two embedding vectors being a proxy for semantic similarity.} using OpenAI's pre-trained \ttt{text-embedding-3-small} model.
\begin{itemize}[leftmargin=*]
\item \textbf{Dataset Embeddings:} Using the dataset title, header, and three example rows as embedding inputs, to support semantic dataset search (\textbf{DC1}). 
\item \textbf{Attribute Embeddings:} Using the column name and the first 10 non-null values as embedding inputs, to support attribute-level filtering (\textbf{DC2}). 
\item \textbf{Dataset Purpose Embeddings:} Using the previously generated dataset description summary and list of purposes as embedding inputs, to support proactive query reformulations (\textbf{DC4}).
\end{itemize}
We stored the augmented and pre-processed dataset collection with all generated embeddings in a PostgreSQL database. We created two HNSW indexes~\cite{malkov2018efficient}: {\em (i)} a \textbf{Dataset Index} using the dataset embeddings (\textbf{DC1}); and {\em (ii)} an \textbf{Attribute Index} using the attribute embeddings (\textbf{DC2}), using the open-source library \ttt{hnswlib}.\footnote{\url{https://github.com/nmslib/hnswlib} (with \ttt{m=16} and \ttt{ef\_construction=64})} Here, given a dataset schema (or an attribute name), the dataset (or attribute) HNSW index returns $k$ most semantically similar datasets (or attributes).

\begin{table}[t]
\centering
\caption{Offline data collection with downstream uses.}
\label{tab:offline-computations}
\scalebox{0.8}{
\begin{tabular}{p{3.6cm} | p{6.4cm}}
\toprule
\textbf{Collected Metadata} & \textbf{Used For} \\
\midrule
• Title + filename + tags & Dataset Cards (Fig. \ref{fig:interface}H) \\
• Dataset Size & Dataset Cards (Fig. \ref{fig:interface}H) \\
• Number of downloads & Dataset Cards (Fig. \ref{fig:interface}H) \\
• Dataset Description & \colorbox{E3Color}{Dataset Embeddings}, Dataset Cards (Fig. \ref{fig:interface}H)\\
• Dataset Sample (10 rows) & \colorbox{E3Color}{Dataset Embeddings}, \colorbox{E2Color}{Attribute Embeddings}, Dataset Cards (Fig. \ref{fig:interface}H)\\
\midrule
\textbf{Generated Metadata} & \textbf{Used For} \\
\midrule
• Description summaries & \colorbox{E1Color}{Purpose Embeddings}, Dataset Cards (Fig. \ref{fig:interface}H) \\
• Attribute descriptions & \colorbox{E2Color}{Attribute Embeddings}, Dataset Cards (Fig. \ref{fig:interface}H) \\
• Data source/collection & Dataset Cards (Fig. \ref{fig:interface}H) \\
• Granularity tags & Granularity Filters (Fig. \ref{fig:interface}F), Dataset Cards (Fig. \ref{fig:interface}H) \\
• Dataset purposes & \colorbox{E1Color}{Purpose Embeddings} \\
\midrule
\textbf{Precomputed Values} & \textbf{Used For} \\
\midrule
• \colorbox{E3Color}{Dataset Embeddings} & Dataset Index for semantic dataset search (Fig. \ref{fig:interface}A) \\
• \colorbox{E2Color}{Attribute Embeddings} & Attribute Index for search \& filtering (Fig. \ref{fig:interface}D, E) \\
• \colorbox{E1Color}{Purpose Embeddings} & Query reformulation suggestions (Fig. \ref{fig:interface}B) \\
\bottomrule
\end{tabular}}
\vspace{-1.5em}
\end{table}

\subsection{Semantic Dataset Search Engine}
\label{sec:system-impl-search}
\tool{} leverages the search indexes (\Cref{sec:system-impl-data-collection} \& \Cref{fig:framework}) to support semantic dataset search (\textbf{DC1}).~\Cref{fig:online-workflow} details the search framework and actions triggered by \tool{} to proactively assist users. The search process begins with users specifying a search query---which may be as brief as a set of keywords, or as detailed as 2--3 sentences. \tool{} uses this query to prompt \ttt{GPT-4o-mini} to generate three diverse hypothetical schemas for a target dataset that would help with the user's query (prompt detailed in~\Cref{appendix:prompts-online}). The generated outputs include the dataset name, projected column names and types, and an example row. These hypothetical schemas capture different ways in which the user’s intent might align with datasets in our collection. Each of the three generated schemas is then embedded using the \ttt{text-embedding-3-small} model, ensuring consistency with previously computed dataset embeddings (\Cref{sec:system-impl-data-collection}). To determine relevance, we compute the cosine similarity between each hypothetical dataset embedding and precomputed dataset embedding pair. Since each of the hypothetical schemas may highlight different aspects of the user's search query, we average the similarity scores obtained for each dataset in our collection for an aggregate similarity score. The datasets are then ranked based on this aggregate score to present the most semantically relevant results. Increasing the number of hypothetical schemas would increase the chances of retrieving highly relevant matches by covering a broader semantic space, but also increase computational costs and query latency. We generate three schemas to balance retrieval effectiveness and response time.

\subsection{Supporting Dynamic and Contextualized Assistance}
\label{system-impl-online}
\tool{} aims to leverage the semantic abilities of LLMs to facilitate contextualized dataset discovery.~\Cref{fig:online-workflow} highlights \tool's online assistance features, and the following sections provide corresponding implementation details.  

\subsubsection{Query Reformulation Suggestions} 
To support \textbf{DC4}, \tool{} surfaces query reformulation suggestions to bridge the gap between user specified dataset search queries and the search space of available datasets (\Cref{fig:walkthrough}B). To do so, \tool{} proactively analyzes all initial dataset search results---performing k-means clustering ($k=15$) over the dataset purpose embeddings (described in~\Cref{sec:system-impl-data-collection}) belonging to the surfaced results---semantically grouping datasets that cover similar topics or have similar intended purposes. \tool{} then picks three clusters that are most relevant to the original search query, and uses an LLM to surface three corresponding query reformulation suggestions (prompt detailed in~\Cref{appendix:prompts-online}), e.g.,~\Cref{fig:walkthrough}B shows the query reformulation suggestion ``analyze the impact of the pandemic on remote work and work-life balance.'' Users may select a query reformulation suggestion to narrow the search scope, or to increase alignment with underlying datasets. Selecting a suggestion refreshes the dataset search results. 

\subsubsection{Semantic Attribute Search and Filter Suggestions}
\tool{} introduces two unique affordances---enabling users to search and filter dataset results based on attribute semantics, instead of exact or fuzzy string matching with attribute names (\textbf{DC2}). First, \tool{} gives users the agency to search by attributes (\Cref{fig:interface}D)---by retrieving relevant datasets based on the HNSW attribute index. That is, given an attribute name, $k$ related attributes from the index are retrieved, and their corresponding datasets are returned, e.g., searching for ``movie name'' will return all datasets containing attributes semantically equivalent to movie titles. Second, \tool{} proactively suggests five ``column concepts'' as filters---informed by both the dataset search results, as well as the user's search query---to narrow down the search space. To do so, \tool{} performs k-means clustering ($k=15$) over the attribute embeddings (described in~\Cref{sec:system-impl-data-collection}) belonging to the datasets in surfaced results and grouping together semantically equivalent attributes. \tool{} then computes a mean vector for each embedding cluster, and computes its cosine similarity with the user's search query. Finally, \tool{} leverages LLM assistance to assign a concept name to the five most relevant attribute clusters, and surface these as filter suggestions (prompt detailed in~\Cref{appendix:prompts-online}), e.g., \ttt{[stress, hours, vacations, employment, remote]} (shown in~\Cref{fig:walkthrough}C).

With these approaches, users may effectively isolate datasets matching attribute-level specifications even if their search terms do not exactly match with column names in a given dataset (\textbf{DC2}).

\subsubsection{Semantic Granularity Filter Suggestions}
As detailed in~\Cref{sec:system-impl-data-collection}, we augmented our collection of datasets with LLM annotations on temporal (e.g. second, minute, hour, ..., year) and spatial (e.g. latitude/longitude, street address, zipcode, ..., country) granularity (\textbf{DC2}). \tool{} also proactively inspects search results to recommend the three most frequently seen \textbf{temporal} and \textbf{spatial} granularity tags as filters (\Cref{fig:walkthrough}D). Users may select a filter to view datasets at the required resolution and level of detail.

\subsubsection{Dynamic Dataset Relevance Indicators}
To assist users in assessing dataset suitability, \tool{} uses LLM assistance to provide in-situ relevance feedback by generating dynamic explanations for dataset utilities and limitations on-the-fly (\Cref{fig:walkthrough}E). To do so, \tool{} considers the user's search query and applied filters, and leverages LLM assistance to generate utility and limitation indicators for the top-5 search results (prompt detailed in~\Cref{appendix:prompts-online}); while relying on lazy-evaluation for the remaining search results, i.e., generating the relevance feedback only if the user clicks on the dataset search result for further inspection. Once generated, all relevance indicators are persisted for future visits to a dataset, unless the user modifies their search query or applied filters. 
\section{User Study: Methodology}
\label{sec:methodology}
To understand how users might leverage \tool's proactive assistance, we conducted a within-subjects repeated-measures study with \numeval{} participants. Our study was guided by the following research questions:

\begin{enumerate}[label={(RQ\arabic*)}, nosep]
    \item How do \tool's features guide users to discover their target datasets? (\Cref{sec:findings-strategies})
    \item How do \tool's capabilities support users' data discovery and sensemaking workflows? (\Cref{sec:findings-gleaning-feedback}) 
\end{enumerate}

\begin{table}[t]
\caption{Participant background and study tasks.}
\label{table:eval-participants}
\small
\scalebox{0.9}{
\begin{tabular}{p{0.3cm} p{0.9cm} p{2.8cm} p{4cm}}
\toprule
\textbf{ID} & \textbf{Order} & \textbf{Background} & \textbf{Tasks} \\
\midrule
P1 & B-C-A & Data Provenance & Neighborhood Migrations in the US \\
P2 & B-A-C & (F4) Art \& AI & Art History and Provenance Data \\
P3 & A-C-B & Databases Researcher & Fraud Detection via ITR \\
P4 & C-B-A & Data Scientist & Question-Answering for LLM-Eval \\
P5 & A-C-B & Data Analyst & Smart-location Sensor Streams \\
P6 & A-B-C & Data Science Graduate & Entity Resolution for Categoricals \\
P7 & C-B-A & (F8) Marine Scientist & Land use for Clean Energy \\
P8 & C-A-B & (F2) AI/ML Engineering & Game Actions Data for Emulations \\
P9 & C-A-B & Business Analyst & Business News Pre-training Data \\
P10 & B-C-A & (F5) ML Engineering & Populating data lake w/ restaurants \\
P11 & B-A-C & Software Developer & Top rated movies and TV Shows \\
P12 & A-B-C & Finance Data Analyst & Financial Inclusion Indicators \\
\bottomrule
\end{tabular}}
\end{table}

\subsubsection*{Recruitment} We recruited \numeval{} participants by emailing prior formative study participants, and through a mailing list of data science professionals maintained by our research group. Four participants overlapped with our formative study (F2 as P8, F4 as P2, F5 as P10, and F8 as P7). All participants had expertise in data science and analytics. They voluntarily consented to taking part in the study, and agreed to have the sessions recorded for transcription and analysis.\del{ Participants were asked to provide details on a search task they would like to perform during the study in the sign-up form. Table 3 reports participants' backgrounds and their self-chosen study tasks.} \add{To maintain ecological validity, participants were asked to bring an open-ended dataset search task of personal relevance, reported in ~\Cref{table:eval-participants}. Participants used the same task across all study conditions to allow for consistent comparisons \cite{hearst2009search}}.

\subsubsection*{Procedure} We conducted a within-subjects repeated-measurements study \add{to facilitate direct comparisons across} three conditions:
\begin{enumerate} [label={(\Alph*)}, nosep]
\item \textbf{Kaggle Dataset Search}: Baseline supporting keyword search (\Cref{appendix:screenshots},~\Cref{fig:kaggle})---chosen for being representative of traditional keyword dataset search tools, as well as for providing a relatively direct comparison standpoint---as \tool's dataset collection is derived from Kaggle;
\item \textbf{Semantic Baseline}: A stripped-down version of \tool{} supporting only semantic search and static metadata filters (\Cref{appendix:screenshots},~\Cref{fig:baseline}), chosen for an experience representative of semantic search tools like Google Dataset Search and Olio's semantic dataset retrieval~\cite{setlur2023olio}; and
\item \textbf{\tool{}}: Complete version with semantic search, query reformulations, filtering, and relevance indicators (\Cref{fig:interface}). 
\end{enumerate}
\del{Participants completed their dataset search task using all three conditions in a randomized order to take experiential learning effects into consideration. We recorded two observations for each ordering.}
\add{Conditions were presented in a randomized order to mitigate learning effects. Participants were also reminded of their original task before each condition to help re-anchor their search, and minimize task drift or carryover from prior conditions and experiences. While participants used the same dataset search task (of their choosing) across conditions, there were no fixed ``correct'' target datasets to be found. Dataset search, like exploratory data analysis, is inherently open-ended~\cite{greenberg2008usability, russell1993cost}. Participants pursued different exploratory trajectories depending on the condition and its affordances, and a dataset search was considered successful if the participant identified one or more datasets to pass their initial round of inspections, and judged them as promising for further investigation.} 
The study began with a brief round of introductions and demographic questions. Each session lasted 60 minutes, during which participants spent 15–18 minutes per condition. Participants were encouraged to think aloud. After each condition, we asked follow-up questions to assess the perceived ease of use of the interface and the relevance of the search results. \add{Participants remotely accessed and controlled a MacBook equipped with an Apple M2 chip, 8GB RAM, and a 10-core CPU. We found the average latency to retrieve datasets to be 1.6 seconds. Suggestions for proactive assistance streamed into the interface within up to 12 seconds.}

Since our system indexed 6,500 datasets from over 50,000 public datasets on Kaggle, we wanted to ensure that participants are not severely restricted by our subset of most popular datasets. To ensure that the semantic baseline and \tool{} had access to relevant datasets, we augmented our initial dataset collection by indexing 300 additional datasets, containing top 25 Kaggle dataset search results for each participant's task. All participants were informed of this dataset scope. To avoid biasing participants, no system walkthrough or tutorial was provided---enabling us to glean their raw impressions and organic usage patterns. This study was approved by our Institutional Review Board (IRB).

\subsubsection*{Analysis}
We transcribed all sessions using Zoom's automatic transcription and supplemented them with detailed notes documenting participant actions throughout the sessions. Two authors performed reflexive thematic analysis through open coding of the transcripts, notes, and screen recordings, followed axial coding to surface broader themes. The authors subsequently performed a second iteration of axial coding to further refine the themes, and achieve high inter-rater agreement. We identified 22 open-codes and 9 axial-codes. \add{Additionally, we analyzed the logs for learning effects across study conditions, and highlight emergent patterns in our study findings.}

\section{User Study: Findings}
\label{sec:user-study}

\thickmuskip=0mu

\begin{table*}[t]
\caption{\add{Task performance and subjective ratings (5-point Likert scale) across study conditions.}
}
\label{table:likert-scale}
\small
\scalebox{0.88}{
\begin{tabular}{
  >{\arraybackslash}m{2.5cm} 
  >{\centering\arraybackslash}m{1.8cm} 
  >{\centering\arraybackslash}m{1.8cm} 
  >{\centering\arraybackslash}m{1.8cm} 
  >{\centering\arraybackslash}m{2.6cm} 
  >{\centering\arraybackslash}m{2.6cm} 
  >{\centering\arraybackslash}m{2.4cm} 
  >{\centering\arraybackslash}m{1.6cm}
}
\toprule
\textbf{Condition} & \textbf{Ease-of-use Ratings$^1$} & \textbf{Relevance Ratings$^2$} & \textbf{\add{\# Queries}} & \textbf{\add{\# Datasets Explored}} & \textbf{\add{Time to assess suitability (s)}} & \textbf{\add{Time to first target (mins)}} & \textbf{\# Successes} \\
\midrule

(A) Kaggle & $\mu=3.08$; $\sigma=0.51$ & $\mu=3.25$; $\sigma=1.05$ & $\mu=3.5$; $\sigma=3.6$ & $\mu=3.33$; $\sigma=1.44$ & $\mu=134$; $\sigma=47$ & $\mu=7.0$; $\sigma=5.4$ & 7 {\em of} 12\\
(B) Semantic Baseline & $\mu=3.75$; $\sigma=0.45$ & $\mu=3.25$; $\sigma=0.86$ & $\mu=1.9$; $\sigma=0.6$ & $\mu=4.25$; $\sigma=1.5$ & $\mu=115$; $\sigma=28$ & $\mu=7.5$; $\sigma=5.5$ & 6 {\em of} 12\\
(C) \tool{} & $\mu=4.75$; $\sigma=0.45$ & $\mu=3.67$; $\sigma=0.78$ & $\mu=1.8$; $\sigma=3.4$ & $\mu=6.02$; $\sigma=2.46$ & $\mu=37$; $\sigma=12$ & $\mu=5.1$; $\sigma=1.7$ & 10 {\em of} 12\\

\bottomrule
\end{tabular}
}
\begin{flushleft}
$^{1,2}$\add{A Friedman test revealed significant differences in ease-of-use ratings across conditions ($\chi^2=23.13$, $p<0.00001$), but no significant differences in relevance ratings ($\chi^2=4.42$, $p=0.11$). Pairwise Wilcoxon tests showed that all ease-of-use comparisons were significant ($p<0.002$, C > B > A). For relevance, only a marginal difference was observed between conditions B and C ($p=0.047$, C > B).}
\end{flushleft}

\end{table*}

Here, we discuss our findings from observing participants engage in dataset discovery workflows across study conditions.

All participants ($n=\numeval$) found \tool{}'s interface to be more {\em ``expressive''} and {\em ``flexible''}, giving them a {\em ``greater sense of control''} over their search task. They appreciated the description summaries and consolidated single-page view---reducing context-switching and scrolling. \textbf{Participants rated \tool{} highly on the ease of use of the interface on a 5-point Likert scale ($\mu=4.75,\sigma=0.45$), and were mostly satisfied with the relevance of search results ($\mu=3.67,\sigma=0.78$)} (\Cref{table:likert-scale}). On the other hand, while using Kaggle, participants echoed sentiments in-line with our formative study findings---being unable to freely express their dataset search intents, finding it restrictive (P2--P4, P6, P10). 
\del{On a 5-point Likert scale, participants expressed mild liking for the baseline interfaces and their search result relevance. Participants had varied success across conditions (Table 4).} \add{\tool also enabled more efficient exploration: participants explored more datasets ($\mu$=6.02), and spent less time in assessing dataset suitability ($\mu$=37s). They also found relevant datasets sooner ($\mu=$5.1 mins). Overall task success was highest with \tool (10 of 12 participants found a relevant dataset), compared to Kaggle (7 of 12) and semantic baseline (6 of 12) (\Cref{table:likert-scale})}.

\add{Across all study sessions, participants used \tool{}'s query reformulation suggestions 15 times (11 of 12 participants), search and filter through column concepts 30 times (12 of 12 participants), and data-granularity filters 3 times (2 of 12 participants).}
We also observed differences in the perceived usefulness of \tool's features to be dependent on the order in which participants were exposed to the conditions. When exposed to \tool before either of the baselines, participants missed the presence of semantic attribute filters the most (P3, P4, P7, P8, P10)---which is the most used feature across sessions (30 invocations); and when exposed to \tool after the baselines, they appreciated the presence of task-specific relevance indicators the most (P2, P6, P11, P12)---which significantly expedited participants' sensemaking and relevance judgments. \add{To examine whether exposure to different conditions influenced user behavior, we analyzed session logs for signs of learning effects. We found that users did not fixate on previously discovered successful datasets; instead, they continued to explore and identify new ones. Notably, when users experienced the control conditions (A or B) first and then transitioned to \tool{} (C), they discovered 12 unique, unseen target datasets (9 of 12 participants). Conversely, when users started with \tool{} and moved to A or B, they still uncovered 7 unseen target datasets (6 of 12 participants).}

On the other hand, we observed differences in dataset search workflows across conditions. First, participants wrote longer and more expressive queries with both \tool{} and the semantic baseline. For example, P2 searched for {\em ``images that are artworks with the names of the artists''} on \tool{}, versus a shorter {\em ``art history''} on Kaggle. 
Second, Kaggle often returned overly selective results (5--20 results), while the semantic baseline returned too many loosely relevant ones (50--100 results). In contrast, \tool{} helped participants start broad with 50+ dataset results, and narrow down to 10--12 datasets effectively using semantic filters, supporting both exploratory and targeted dataset search workflows.
Lastly, participants frequently downloaded datasets in the baseline conditions for deeper inspection. With \tool{}, this need diminished due to in-situ feedback from relevance indicators. \add{All participants noted the usefulness of such indicators, and 8 of 12 commented on their soundness and credibility.}

In what follows, we present qualitative findings from the user study, organizing them around two key capabilities \tool{} unlocked for users: first, their ability to steer and refine their search through interactive features (addressing RQ1); and second, their ability to adapt to search results and learn during exploration (addressing RQ2).

\subsection{\tool{} Unlocked Users' Ability to Steer and Adapt Their Dataset Search}
\label{sec:findings-strategies}

\tool{} enabled participants to adopt more deliberate and informed dataset search strategies (P1, P4, P5, P7, P8, P10, P12). Compared to the baselines, users learned to steer system feedback to their advantage (P2, P3, P6, P8, P10), and encountered learning moments that enhanced their sensemaking and search behavior—even beyond \tool's immediate environment (P4, P6, P8, P9). We describe these distinctive strategies below.

\subsubsection{Users learned to ``prompt-engineer'' queries to control \tool{}'s relevance indicators}
\label{subsec:knobs-relevance}

Participants learned through interaction that the dimensions of feedback highlighted by the relevance indicators was dependent on their query and filters (P1--P3, P6, P8--P12). As they gained increased familiarity with \tool{}, some participants began treating their queries as ``knobs'' they could use to manipulate the dataset relevance indicators (P2, P3, P6, P8--P10)---adjusting their task descriptions to elicit more targeted and informative feedback from the system. For instance, P2 needed information about image use rights for datasets containing links to artwork images. They hypothesized that modifying the query with this request would affect the relevance indicators, and added---``I need to know what the image rights are (e.g. if it is public domain, CC0, if attribution is required, etc.).'' Thereafter, the relevance indicators began surfacing image licensing details for each dataset.

Similarly, P3 mentioned their preference for ``non-synthetic'' datasets in their query---with the objective of having relevance indicators pin-point dataset sources upfront.
This contrasts with our formative study findings, where participants held unspoken dataset relevance criteria and felt restricted by the dataset search interfaces.
By making relevance indicators visible and responsive, \tool{} successfully elicited hidden preferences---promoting a reflective search process for other participants as well (P6, P8--P10).

\subsubsection{\tool{} empowered users by enabling fine-grained queries over dataset attributes and granularity levels}
Participants used \tool's features (query reformulations, and attribute and granularity filters) to systematically broaden or refine their search (P1, P4, P5, P7, P8, P10, P12). P7 began with the query: ``land use in USA,'' which returned mostly irrelevant results, and then used \tool's query reformulation suggestion---``land distribution across countries''--- to consciously broaden the scope. This surfaced more relevant, but geographically non-localized datasets. With this broader scope, \tool also suggested the \ttt{country-level} granularity filter, enabling P7 to narrow results back down to the desired resolution, albeit requiring some pre-processing to filter out all non-U.S. records. This tandem-use of query reformulation suggestions and semantic granularity filters exemplifies how \tool supports exploration followed by targeted narrowing. We observed similar workflows with \tool supporting concerted refinement efforts for P1, P5, P7, P10, and P12. Notably, each of these participants had embarked on discovering geographical data with varied levels of granularity.  

Through using \tool's semantic attribute search, participants were able to not only narrow down the search space, but also stumble across previously latent datasets (P1, P2, P12). For instance, P2 had been deeply invested in their search for art history datasets prior to our evaluation study, and described extensively using Kaggle for this task. P2 used the semantic attribute search---a new dataset search modality surfaced by \tool{}---to intentionally look for datasets with the \ttt{"artist bio"} column, leading them to discover a previously unknown dataset (Carnegie Museum Collections) that was highly relevant to their work. They appreciated the system's semantic matching, noting, {\em ``it’s great that it is not only exact matching the column name but it gets the vibes.''} We observe how \tool can surface useful datasets even for other experienced participants working in familiar domains (P1, P12).

\subsection{\tool Helped Users Make Sense of Dataset Availability}
\label{sec:findings-gleaning-feedback}

Participants frequently repurposed \tool's features to gain feedback on their queries (P1, P3, P4, P7, P8, P10, P12), build conceptual models of the search space (P4, P9, P10, P12), and sanity-check their progress (P2, P5, P7, P8, P12). Users actively interpreted \tool's proactive reformulation and semantic filtering suggestions---turning them into implicit system feedback to reason about dataset availability, recalibrate expectations, and steer their search strategy.

\subsubsection{Relevance indicators triggered ``aha'' moments that changed how users judged datasets}
Beyond immediate task success, \tool{} prompted meaningful learning moments that shaped users’ dataset suitability assessment strategies.
For some participants, learning moments emerged as a byproduct of expediting sensemaking through dataset relevance indicators, making connections or limitations apparent upfront.
For example, P6 initially dismissed a dataset surfaced by the semantic baseline as irrelevant. However, when the same dataset appeared in \tool, they reviewed the system's utility explanation and reconsidered its fit. The system had highlighted `joinable' columns relevant to P6's knowledge graph task, helping them realize the applicability of the dataset.
P6 noted, {\em ``it provides reasoning and is quite responsive... it [utility indicators] helped me understand what to expect from the dataset.''} This illustrates how transparent, in-context explanations can change user perceptions. P6 then continued looking for datasets with a renewed lens for dataset applicability. We observed similar patterns with P4 and P9. Notably, each of these participants' tasks were geared towards finding datasets that would serve as inputs to algorithms they have authored themselves---offering some flexibility in how the dataset or their algorithm can be adapted to each other.

Interestingly, for one participant (P8), the LLM generated relevance indicators enabled a learning moment by filling an information retrieval need. P8 began with a clear objective: ``predicting NBA game outcomes based on LaMello Ball's three-point shots.''  While reviewing a dataset from 2008--2014, \tool's relevance indicators surfaced a limitation: \ttt{``LaMelo started playing for Charlotte Hornets in 2020, while the time-span of this dataset predates LaMelo's NBA career.''} This insight helped P8 quickly rule out the dataset and refine their assessment criteria for the remainder of the study---while carrying this learning over to Kaggle, where they began checking dataset upload dates more deliberately. 

\subsubsection{Users adapted their queries when query reformulation suggestions hinted at unavailable data}
Participants learned early on that the query reformulation suggestions were dependent on the search results yielded by \tool (P1, P3, P4, P6--P8, P10, P11). Some used these suggestions to verify whether their queries contained enough detail (P1, P7), while others used them to make bets on the presence of relevant datasets, probe the search space, and adapt their expectations (P3, P4, P8).

For instance, P3 originally searched for non-synthetic money transfer datasets on Kaggle. However, \tool and baseline did not have any real-world money transfer datasets as part of their dataset collection, leading to irrelevant results based on synthetic sources. This mismatch led them to question the reliability of the results: {\em ``I started to lose faith in the results and their ranking''}. However, the reformulation suggestion \ttt{“Analyze anomalies in real-world income tax datasets”} hinted at not only the absence of money transfer datasets, but the abundance of real-world income tax anomaly datasets; helping P3 pivot their task to income tax datasets---realigning their goals to match the available search space. Other participants refined their geographic or demographic focus without changing their broader goals. For example, P12 used reformulation suggestions to scope financial inclusion data down to agricultural workers in Rwanda.


Relevance indicators also played a role in helping participants evaluate the viability of their queries (P4, P9, P10, P12). When one or more top-ranked datasets indicated \ttt{"No significant utilities"} (highly ranked datasets showing poor task adherence)---prompted participants to reformulate their queries.\footnote{While participants in our formative study also encountered irrelevant top-ranked results in using semantic dataset search engines (like Google Dataset Search), they typically skipped to the next entry without reflecting on the mismatch between ranking and task relevance. We believe that \tool{}’s relevance indicators prompted users to re-express intent, enabling more iterative and reflective searching. We hypothesize that the presence of relevance indicators but facilitate \textit{\textbf{meta-cognition}}---helping users reason not only about what they see, but also about their next steps, as discussed in the Cognitive Fit theory by ~\citet{vessey1991cognitive}.} On facing this conflict, P10 said, {\em ``No significant utilities higher up in the search results means that I should change my query, seems like there is not a lot in the search space to begin with.''}

\subsubsection{Seeing the ``right'' semantic filter suggestions gave users confidence they were on track} 
Participants also experientially learned that the suggested semantic attribute and granularity filters depended on the search results (P2, P5, P7, P8, P12). Over time, these filter suggestions became feedback signals or sanity checks that participants used to validate their current direction.
Seeing the {\em ``right''} filter suggestions reassured participants that they were on the right track, and within their intended space of dataset search results. For instance, P12 noted, {\em ``Seeing \ttt{[agriculture, income, credit]} is affirmative of my intent---it tells me I am still in the right space.''} In contrast, when filter suggestions seemed off, participants interpreted that as a sign to revise their query. P5, searching for ``intergenerational facilities,'' initially saw unrelated filters like \ttt{[emissions, source, insurance, url]}, prompting them to rethink their query phrasing. After revising the query, more aligned filters appeared, such as \ttt{[daycare, address, age, cost]}, reinforcing their revised direction.

 Similarly, P7 said, {\em ``I see emissions, energy, land, population, and water, along with a year-level filter suggestion. This is giving me confidence that your system is understanding my prompt correctly.''} P8 also supported our observation, mentioning how these acted as early cues: {\em ``even before I look at the search results, the smart column filters are giving me some clue about the kind of data in the search results.''} \tool's semantic filters suggestions served as both, conceptual scaffolds, and lightweight progress markers during open-ended search tasks.
\section{Discussion}
We reflect on our findings in context of sensemaking and information-seeking literature, and discuss opportunities to extend \tool. 

 \subsection{Impact of Relevance Indicators on Sensemaking} 
Our findings show how \tool supported sensemaking through \del{query reformulation suggestions (helping users reason about the search space; see Section 7.2), and } relevance indicators, helping users assess dataset suitability (see~\Cref{sec:findings-gleaning-feedback}). We interpret these findings through~\citet{kaur2022sensible}'s framework on sensible AI explanations, which emphasizes understanding not just the content of explanations, but their cognitive timing and alignment with user goals as well.

{\em Relevance indicators support \textbf{Identity Construction} by affirming users' intents.}
Relevance indicators helped users quickly identify datasets that aligned with their stated goals and preferences. Echoing~\citet{kaur2022sensible}, we found that participants gravitated towards cues that affirmed their own reasoning---using them to either confidently shortlist datasets, or skip them without further inspection---speeding up their workflow \add{(as seen in~\Cref{table:likert-scale})}.

{\em Relevance indicators disrupt \textbf{Retrospective Sensemaking}.}
\citet{kaur2022sensible}'s framework argues that offering explanations before users have had a chance to reflect on information themselves negatively affects their sensemaking. In our case, \tool immediately surfaces task-specific relevance indicators upon inspecting a dataset---often leading to quick decision-making. \del{Recall that we observed how participants no longer expressed the need to download datasets for further inspection while using DataScout. }Surfacing such cues too early sometimes disrupted users’ independent judgment of dataset suitability, and short-circuited their exploratory and sensemaking processes. Complementary to this argument, P2 and P5 voiced concerns about the subjectivity in LLM interpretations, preferring to view the {\em ``raw data''} and {\em ``hard cold facts,''} over {\em ``narratives around the data.''} This skepticism echoes prior work on interactive ML systems, where~\citet{groce2013you} observed users heavy reliance on visible system cues while remaining wary of subjective or opaque feedback.

Yet, users also wanted more visible and persistent indicators (P1, P8).\footnote{P8 suggested a simple, persistent thumbs-up/down mechanism; and P1 wanted always-on relevance indicators to avoid clicking on each dataset for further inspection.} These opposing reactions reflect a fundamental tension: if surfaced too early, sensemaking aids can overly steer users; if surfaced too late, they may lose their utility altogether; as also discussed by~\citet{amershi2019guidelines}. Future dataset search systems must negotiate this tradeoff carefully, perhaps by layering relevance signals across interaction stages and interface elements.

\del{In contrast, DataScout's query reformulation suggestions facilitated the kind of retrospective sensemaking that Kaur et al. [24] advocate. Rather than steering users prematurely, these suggestions supported reflective reasoning—helping participants assess whether their query was viable in the given dataset search space. As such, query suggestions acted not just as reformulations, but as collaborators in helping users understand and realign their search strategies.}

\subsection{Operationalizing Structured Exploration}
While \tool{} supports dataset discovery through NL intent expression, participants expressed a need for more structured control over their query's interpretation---specifically, the ability to specify binary constraints in NL, rather than loose preferences. This reflects a common tension in semantic search: while NL offers flexible intent expression, it can \del{obscure whether specified criteria are being treated as strict filters or loose preferences.} \add{blur the line between strict filters and preferences, limiting users' ability to precisely steer their search.}
\add{Participants envisioned interfaces to distinguish between constraints and preferences---}P9 suggested separate input fields for the two, prompting reflection on search goals; while P2 proposed an adaptive \del{system that could treat their query criteria as hard constraints when the result set is too large, and as preferences when it is too sparse.}\add{mechanism that can automatically treat criteria as constraints when results are too broad, and as preferences when too narrow---}mirroring the \textit{\textbf{Information Diet Model}}, where users must balance preferences (easy-to-catch prey) and rigid constraints (hard-to-catch-prey) to optimize search~\cite{pirolli2007IFT}.\footnote{{\em ``If a predator is too specialized, it will do very narrow searching. If the predator is too generalized, then it will pursue too much unprofitable prey''~\cite{pirolli2007IFT}}} 
Prior work in exploratory search\del{ and faceted interfaces} has emphasized\del{ the value of} supporting both fluid and rigid filtering modes as well~\cite{marchionini2006exploratory, hearst2006clustering, rossi2008preferences, hearst2009search}. \del{, and interfaces that allow toggling between these modes have been shown to reduce abandonment [18]. However, limited research has focused on automatically extracting such constraints from natural language inputs [27] or applying them meaningfully in dataset search contexts.}

\del{As dataset search systems evolve, a promising direction is the development of mixed-initiative filters that can clarify how NL specifications are interpreted---allowing users to steer their search more precisely.}

\add{A complementary direction involves expanding \tool{}'s query reformulations beyond their current role of narrowing results. Reformulations could also broaden the search space by introducing adjacent and semantically related results, helping users consider alternatives they may not have explicitly articulated, thus supporting robust exploration through both---structured narrowing and expanding of the dataset search space.}

\subsection{Limitations and Future Work}
Our evaluation of \tool{} has several limitations. First, the search precision was constrained by our collection of Kaggle datasets, occasionally producing irrelevant results despite augmenting our corpus with $\sim$300 datasets for participant tasks. Second, our prototype lacked basic search functionalities such as result sorting and support for varied ranking criteria (e.g. upload date, downloads, size), which limited participants’ ability to explore results systematically. Third, we recorded only two observations per condition order, limiting findings on experiential effects. Finally, we compared only with Kaggle as a keyword-search baseline due to our shared dataset sources, and lacking access to other deployed systems with similar data. This choice allowed for direct comparisons, but narrowed our evaluation scope.

We suggest several directions for extending \tool{}. First, users desired a {\em ``birds-eye view''} (P7) summarizing patterns across results---such as covered time periods or geographic regions---to expedite sensemaking and offer feedback on their queries (\Cref{sec:findings-gleaning-feedback}). Aggregated overviews, as explored by~\citet{ouellette2021ronin}, could support this need by presenting bottom-up hierarchical summaries of results.
Second, users often wanted to combine data from multiple sources to construct their intended dataset (F2, F5, F7, F8)---via union or joins. While prior work has addressed union/join-based dataset search, future interfaces could better support this with tailored sensemaking tools and visual cues for multi-dataset compositions.
Finally, participants wanted visibility into data quality (P4, P10, P12). Building on existing efforts in data quality detection and wrangling~\cite{kandel2011wrangler, guo2011proactive, chopra2023cowrangler, skrub2025}, future systems could surface these cues as relevance indicators to better inform user decisions.
\section{Conclusion}
We introduce \tool{}---a system that rethinks dataset discovery through proactive AI-assistance, offering query reformulation suggestions, semantic search and filtering based on attributes and data granularity, and task-specific dataset relevance indicators---supporting users in navigating and understanding opaque dataset landscapes. Our study with \numeval{} participants revealed how these features expedited sensemaking and conceptual model building; while eliciting latent search specifications. Our findings also underscore the need for dataset search systems to be designed to support both, exploratory wandering and targeted retrieval---meeting users where they are in their evolving dataset search workflows.

\begin{acks}
We are grateful to HC Moore, Yiming Lin, Sepanta Zeighami, James Smith, and Hila Mor for their valuable feedback on our prototypes and findings. We thank our study participants for their engagement and feedback, helping us identify constructive search workflows and unmet needs for dataset search systems. This work was supported by the National Science Foundation (grants DGE-2243822, IIS-2129008, IIS-1940759, and IIS-1940757), the Dutch Research Council (NWO, grant NGF.1607.22.045), funds from the State of California, an NDSEG and BIDS-Accenture Fellowship, funds from the Alfred P. Sloan Foundation, as well as EPIC Lab sponsors (Adobe, Google, G-Research, Microsoft, PromptQL, Sigma Computing, and Snowflake).
\end{acks}
\balance
\bibliographystyle{ACM-Reference-Format}
\bibliography{references}

\newpage

\appendix
\section{Screenshots of User Study Conditions}
\label{appendix:screenshots}

\begin{figure}[htp]
    \centering
    \includegraphics[width=\linewidth]{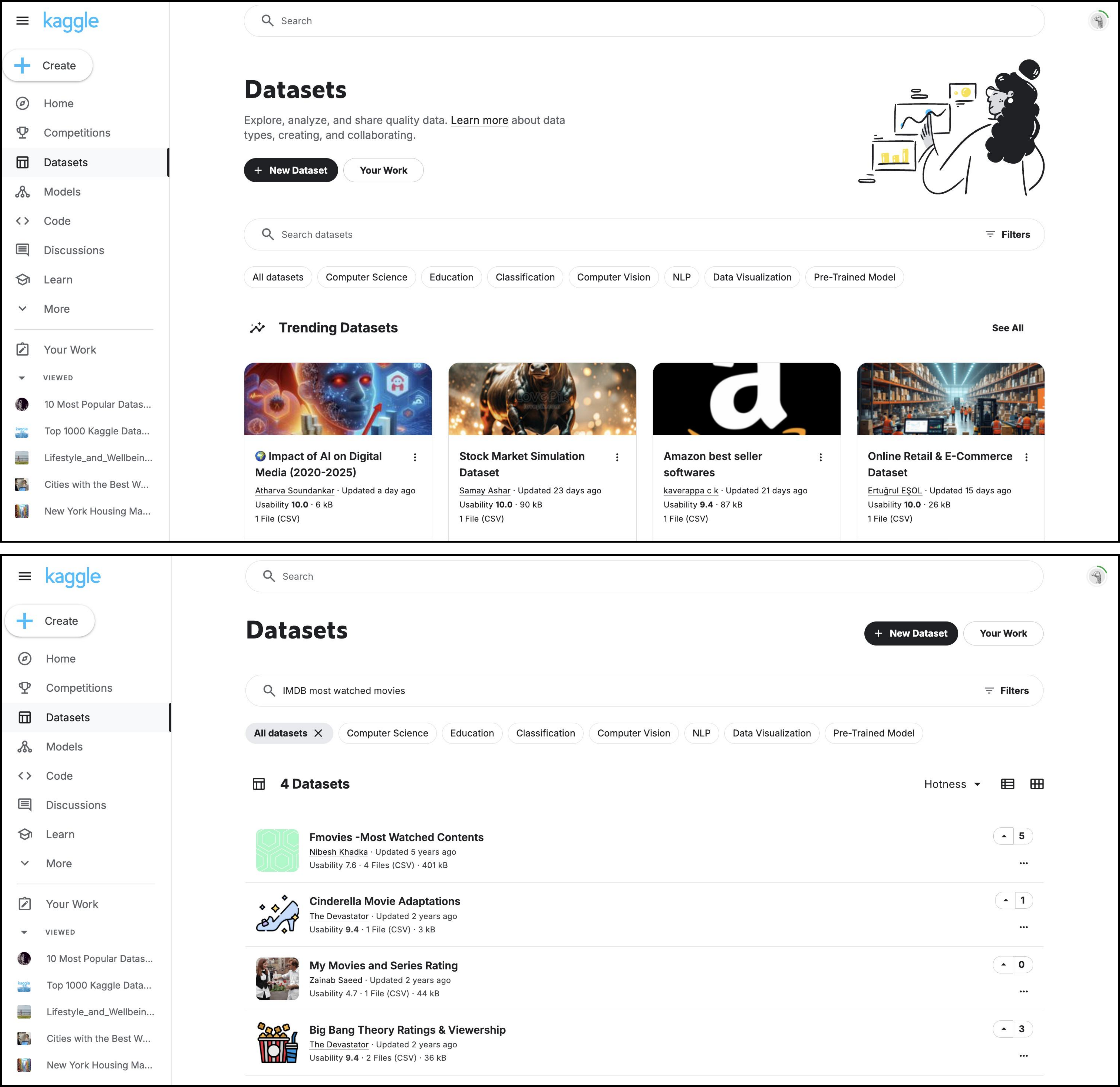}
    \caption{(A) Kaggle: Keyword dataset search condition}
    \label{fig:kaggle}
\end{figure}

\begin{figure}[htp]
    \centering \includegraphics[width=\linewidth]{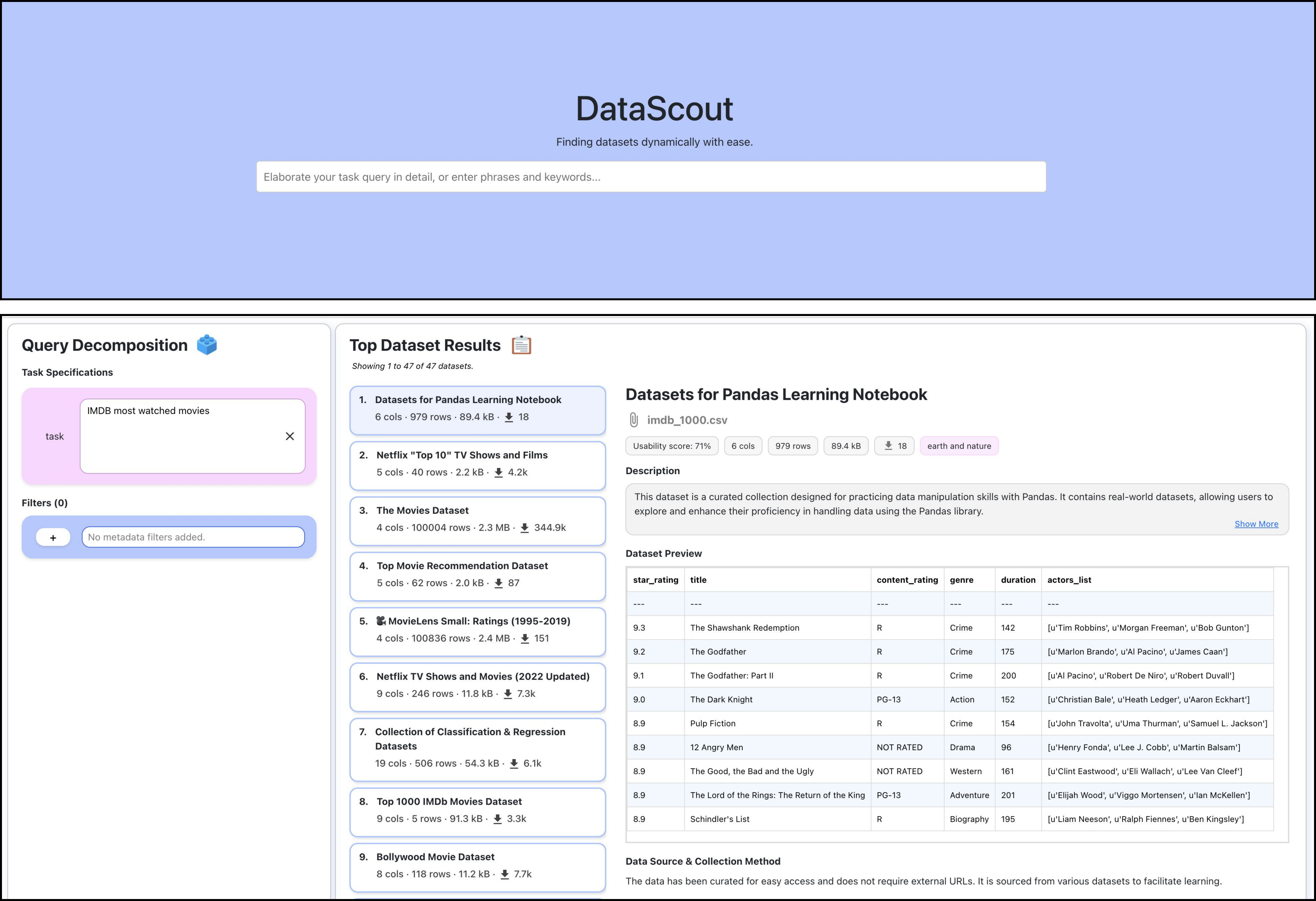}
    \caption{(B) Semantic Baseline: A stripped-down version of \tool{} supporting only semantic dataset search.
    }
    \label{fig:baseline}
\end{figure}

Here, we present screenshots for our baseline dataset search conditions, as detailed in~\Cref{sec:methodology}: (A) Kaggle (\Cref{fig:kaggle}) and (B) Semantic Baseline (\Cref{fig:baseline}).

\section{Prompts for Offline Metadata Augmentation}
\label{appendix:prompts-offline}

We designed two prompts to augment the metadata for each dataset in our collection, as described in~\Cref{sec:system-impl-data-collection} and illustrated in~\Cref{fig:framework}. For every dataset, we prompted OpenAI's \ttt{GPT-4o} to: {\em (i)} enrich the metadata with summaries of the dataset’s description, purpose, source and collection methods, and attribute descriptions; and {\em (ii)} annotate the dataset's temporal and spatial granularity. We use the LLM's tool calling functionality to generate structured responses that adhere to the output schema defined for each prompt.

\focusTextBox{\underline{Dataset Metadata Augmentation} \vspace{0.3em}}
{Given following dataset details, you must extract information about this dataset.\\
\textbf{Dataset Details:}
\begin{itemize}
\item Title: \textbf{\ttt{\{title\}}}
\item Description: \textbf{\ttt{\{description\}}} 
\item Dataset Preview: \textbf{\ttt{\{example\_rows\}}}
\end{itemize}

\noindent Directly answer each question, be brief and to the point: \\
\textbf{1. Description Summary:} In 1--3 sentences, provide a brief and summarized description of the dataset.\\
\textbf{2. Purposes:} Provide a list of analytical, data science, visualization, or machine learning tasks that can be performed with this dataset. e.g., \ttt{["training a regression model", "temporal analysis"]} \\
\textbf{3. Dataset Source \& Collection Methods:} Gather the source(s) of this dataset, which could include names and/or affiliations of persons, website URLs, web-APIs, synthetic sources, human annotations, and so on. If no information is available about the source of the data, output \ttt{`N/A'}. \\
\textbf{4. Column Descriptions:} For each column in the dataset, provide a brief description for the column with its data type. \\
\textbf{Output Schema:} \\
\ttt{\{``description\_summary'': string, \\
``dataset\_purposes'': list[string], \\
``dataset\_sources'': string, \\
``column\_descriptions'': list[\{``column\_name'': string, ``type'': string, ``description'': string\}]
\}}
}

\focusTextBox{\underline{Temporal \& Spatial Granularity Annotation} \vspace{0.3em}}
{Given a dataset with the following details, determine the most likely temporal and/or spatial granularity reflected in the dataset.\\
\textbf{Dataset Details:}
\begin{itemize}
    \item Title: \textbf{\ttt{\{title\}}}
    \item Description: \textbf{\ttt{\{description\}}}
    \item Dataset Preview: \textbf{\ttt{\{example\_rows\}}} 
\end{itemize}

\noindent Select the \textbf{temporal granularity} from the following options: \\
Year, Quarter, Month, Week, Day, Hour, Minute, or Second. \\
Select the \textbf{spatial granularity} from the following options: \\
Continent, Country, State/Province, County/District, City, Neighborhood/Region, Zip Code/Postal Code, Street Address, Residential Address, or Latitude/Longitude. \\
Identify the temporal and/or spatial granularity only if reflected in the dataset. Leave the respective field(s) empty if the granularity cannot be inferred from the table.\\
\textbf{Output Schema:}\\
\ttt{\{``temporal\_granularity'': string, \\
``spatial\_granularity'': string\}}
}

\section{Prompts for Online Dataset Search Assistance}
\label{appendix:prompts-online}

We designed four prompts to support \tool’s semantic search and online LLM-in-the-loop workflows, as detailed in Sections~\ref{sec:system-impl-search} and~\ref{system-impl-online}, and illustrated in~\Cref{fig:online-workflow}. These prompts take the user’s dataset search query, applied filters, and the resulting datasets as inputs, enabling proactive and contextualized assistance. \tool{} sends these prompts to OpenAI’s \ttt{GPT-4o-mini}, once again leveraging its tool calling functionality to generate structured responses that follow the output schema defined for each prompt.

\focusTextBox{\underline{Hypothetical Schema Generation} \vspace{0.3em}}
{Given the task of \textbf{\ttt{\{query\}}}, generate three dataset schemas to implement the task.
Only generate three table schemas, excluding any introductory phrases and focusing exclusively on the tasks themselves.
Generate the table names and corresponding column names, data types, and example rows. For example:\\
\textbf{Example Task:}
Datasets to train a machine learning model to predict housing prices

\noindent \textbf{Example Output:} (Parts omitted for brevity)\\
\ttt{
[
  \{
    "table\_name": "Properties",\\
    "column\_names": ["id", "num\_bedrooms", "num\_bathrooms", \\
    "sqft", "year\_built", "location", "price"],\\
    "data\_types": ["INT", "INT", "INT", "FLOAT", "INT", "TEXT", "FLOAT"],\\
    "example\_row": [101, 3, 2, 1450.5, 2005, "Seattle, WA", 675000.0]
  \},\\
\{
    ``table\_name'': ``NeighborhoodStats'',\\
    ``column\_names'': [...],\\
    ``data\_types'': [...],\\
    ``example\_row'': [...]
  \}, \\
  \{
    ``table\_name'': ``PropertySalesHistory'',\\
    ``column\_names'': [...],\\
    ``data\_types'': [...],\\
    ``example\_row'': [...]
  \}
]}\\

\noindent \textbf{Output Schema:}\\
\ttt{list[
\{"table\_name": string, \\
"column\_names": list[string], \\
"data\_types": list[string], \\
"example\_row": list[string]\} 
]}
}

\focusTextBox{\underline{Generate Query Reformulations} \vspace{0.3em}}
{Generate a dataset search query matching a collection of given dataset names, such that it:
\begin{itemize}[leftmargin=*]
\item Incorporates the common theme of these dataset names: \textbf{\ttt{\{cluster\}}}
\item Relates to the original task: \textbf{\ttt{\{query\}}}
\item Is specific enough to include both a topic, as well as a clear objective.
\end{itemize}

\noindent Also provide a brief reason (under 10 words) why this query improves upon \textbf{\ttt{\{query\}}}.\\
\textbf{Example Output:}\\
\ttt{
\{
    "query": "Analyze voter demographics in presidential elections", "reason": "adds demographic focus"
\}} \\

\noindent \textbf{Output Schema:}\\
\ttt{\{"query": string, "reason": string\}}
}

\focusTextBox{\underline{Generate Column Name Concepts} \vspace{0.3em}}
{You are an assistant that returns a flat list of words. The input will be a list with nested elements. For each nested element, return 1 to 2 representative words that best represent the topic of the nested group. The representative word should also make sense in context with the \textbf{\ttt{\{query\}}}. The words should be lower case single words without special characters (like hyphens or underscores). The output must be a valid JSON array with no additional formatting, symbols, or repetitions.\\

\noindent \textbf{Output Schema:}\\
\ttt{list[string]}
}

\focusTextBox{\underline{Generate Relevance Indicators} \vspace{0.3em}}
{You are an assistant that explains what makes the following dataset search result relevant or irrelevant, given my task and applied search filters.

\noindent \textbf{Dataset Details:}
\begin{itemize}
\item Description: \textbf{\ttt{\{description\}}}
\item Example Rows: \textbf{\ttt{\{schema\}}}
\item Purpose of dataset: \textbf{\ttt{\{purpose\}}}
\item Dataset Collection Method: \textbf{\ttt{\{source\}}}
\end{itemize}

\noindent \textbf{Dataset Search Specifications:}
\begin{itemize}
\item Dataset search query: \textbf{\ttt{\{query\}}}
\item Applied filters: \textbf{\ttt{\{filters\}}}
\end{itemize}

\noindent \textbf{Instructions:}\\
1. Utilities: Identify the strongest factors that make this dataset useful. Look for the presence of relevant attributes, high data quality, and matching intent. If there are no strong advantages, return \ttt{"No significant utilities."}

\noindent 2. Limitations: Identify limitations such as missing relevant attributes, specific geographical locations (e.g., ``dataset only contains records of location X''), specific temporal ranges (e.g., ``data belongs to X and Y time range''), poor data quality and missing or incomplete data. If no major issues exist, return \ttt{"No significant limitations."}

\noindent \textbf{Guidelines:}
\begin{itemize}[leftmargin=*]
\item Stay factual: Base responses strictly on the provided dataset details. Do not assume information that isn’t explicitly stated.
\item Be concise: Limit each response to 1--2 sentences.
\item Avoid hallucination: If no strong reason exists for relevance or irrelevance, default to \ttt{"No significant utilities"} or \ttt{"No significant limitations"}.
\end{itemize}

\noindent \textbf{Output Schema:}\\
\ttt{\{"utilities": string, "limitations": string\}}
}

\end{document}